\documentclass{article}

\begin{document}

\title{{\bf Transition matrix Monte Carlo and flat-histogram algorithm}}

\author{Jian-Sheng Wang\\
Singapore-MIT Alliance and Department of Computational Science,\\
National University of Singapore,\\
Singapore 119260, Republic of Singapore}

\date{17 June 2003}

\maketitle
\begin{abstract}
In any valid Monte Carlo sampling that realizes microcanonical
property we can collect statistics for a transition matrix in energy.
This matrix is used to determine the density of states, from which
most of the thermodynamical averages can be calculated, including free
energy and entropy.  We discuss single-spin-flip algorithms, such as
flat-histogram and equal-hit algorithms, that can be used for
simulations.  The flat-histogram algorithm realizes multicanonical
ensemble.  We demonstrate the use of the method with applications to
Ising model and show its efficiency of search for spin-glass ground
states.
\end{abstract}



\section{Introduction}
    
In traditional Monte Carlo sampling method, the computation of a
thermodynamic quantity $\langle Q \rangle$ is usually through a simple
arithmetic average:
\begin{equation}
\langle Q \rangle = { 1 \over M} \sum_{i=1}^M Q(\sigma_i),
\end{equation}
where the configurations $\sigma_i$ are generated according to a
specified distribution, such as the Boltzmann distribution.  However,
it is possible to collect other information in the same simulation,
from which we can obtain better statistics, or estimates of quantities
other than that at simulation parameters.  The histogram method
\cite{histo1} and multi-histogram method \cite{histo2} collect energy
histogram at a given temperature, from which the quantity at nearby
temperature can be inferred.  The key observation here is that the
histogram of energy is related to density of states through $H(E)
\propto n(E) \exp(-E/kT)$ (in canonical ensemble).  From the
histogram, we can determine the density of states $n(E)$.  Once the
density of states is known, we can compute most of the thermodynamic
quantities at any temperature.

Histogram method has been found to be an excellent tool for study
critical phenomena.  Further improvement can be made by collecting
`high-order' statistics, i.e., the transition matrix
\cite{wang-tay-swendsen-PRL,Li-thesis,JSP}.  With histogram method,
each configuration provides just an `1' to a histogram entry, while in
transition matrix method, each configuration gives several numbers of
magnitude about $N$ to the transition matrix elements, thus variance
reduction is expected.  One of the most appealing feature of
transition matrix Monte Carlo is an easy and straightforward way to
combine several simulations.  Additionally, we can use any valid
sampling algorithm in a generalized ensemble which realizes the
microcanonical property that states with same energy have the same
probability.  The flat-histogram algorithm
\cite{wang-europhys-B,swendsen-int,wang-lee} is such an algorithm that
realizes multicanonical ensemble in which the energy histogram
distribution is a constant.  In the following, we present the
transition matrix Monte Carlo method, introduce the flat-histogram and
other related algorithms.  We discuss the performance of algorithms
with examples from Monte Carlo simulation results of the Ising models.
We summarize in the last section.

\section{Transition Matrix Monte Carlo Method}
First, we give the definition for the transition matrix.  Let
$W(\sigma \to \sigma')$ be the transition probability of the states
from $\sigma$ to $\sigma'$ of a Markov chain.  To be definite, we
consider a single-spin-flip dynamics with a canonical distribution,
but the formalism is general.  The transition matrix in the space of
energy from $E$ to $E'$ is
\begin{equation}
T(E \to E') = {1 \over n(E) } 
              \sum_{E(\sigma)=E} \sum_{E(\sigma')=E'} W(\sigma \to \sigma'),
\end{equation}
where the summations are over all initial states $\sigma$ with energy
$E$ and all final states $\sigma'$ with energy $E'$.  Estimates of the
transition matrix can be obtained during a Monte Carlo sampling, where
the summation over $E$ divided by $n(E)$ is interpreted as a
microcanonical average of the state-space transition probabilities of
the Markov chain, i.e.,
\begin{equation}
T(E \to E') = \sum_{E(\sigma')=E'} 
\bigl\langle W(\sigma \to \sigma') \bigr\rangle_E,
\end{equation}
The expression can be further simplified if we consider
single-spin-flip dynamics, with a spin choosing at random.  In this
case, $\sum_{E(\sigma')=E'}W(\sigma \to \sigma') = {1 \over N}
N(\sigma, \Delta E) a(E \to E')$, where $N(\sigma, \Delta E)$ is the
number of sites such that a spin-flip causes the energy increasing by
$\Delta E = E'-E$ in the current state $\sigma$.  It is also the
number of possible moves that one can make to change the energy by
$\Delta E$.  Note that $\sum_{\Delta E} N(\sigma, \Delta E) = N$,
where $N$ is the number of sites.  A common choice of the
single-spin-flip rate $a(E \to E')$ is the Metropolis rate
$\min\bigl(1, \exp(-\Delta E/kT)\bigr)$.  Since this factor is a
function of $E$ and $E'$, the microcanonical average $\langle \cdots
\rangle_E$ is performed over $N(\sigma, \Delta E)$ only.  We have
\begin{equation}
T(E \to E') = {1 \over N} \bigl\langle N(\sigma, \Delta E)\bigr\rangle_E a(E \to E') 
= T_\infty(E \to E') a(E \to E'),
\label{Teq}
\end{equation}
where we have defined a normalized $N(\sigma, \Delta E)$ as the
infinite temperature transition matrix.

The eigenvector corresponding to the eigenvalue 1 of the transition
matrix is the probability of finding states with energy $E$.  It is
also proportional to the histogram $H(E)$.  To determine the density
of states, a numerically better choice is from the detailed balance.
This gives us the relationship between histogram and transition
matrix:
\begin{equation}
H(E) T(E \to E') = H(E') T(E' \to E).
\end{equation}
If we use the fact that $H(E) \propto n(E) \exp(-E/kT)$ and
Eq.~(\ref{Teq}), we obtain the so-called broad-histogram equation
\cite{oliveira-brasil1,oliveira-europhys-b-2,oliveira-broad,berg-hansmann-europhys-b}
\begin{equation}
n(E) \bigl\langle N(\sigma, E'-E)\bigr\rangle_E = 
   n(E') \bigl\langle N(\sigma', E-E') \bigr\rangle_{E'}.
\label{bheq}
\end{equation}
This is one of the basic equation for determining the density of
states, as well as for the flat-histogram algorithm below.

\section{Flat-Histogram Algorithm}
Any sampling algorithm that can realize microcanonical property, i.e.,
the distribution of the states is a function of energy only, can be
used to collect statistics for $\langle N(\sigma, \Delta E)\rangle_E$.
Using a canonical ensemble simulation, we need dozen temperatures in
order to cover all the relevant energies.  However, comparing to
multi-histogram methods, the combination of data at different
temperatures is very easy, we simply add up the matrix elements and
then properly normalize.

Multicanonical ensemble \cite{berg} is a particularly good choice for
the collection of transition matrix elements, since it reaches all
energy levels with equal probability.  Multicanonical ensemble is
defined to be $H(E) = {\rm const}$, or the probability of
configuration $P(\sigma) \propto 1/n\bigl(E(\sigma)\bigr)$.  It is
purely an artificial ensemble designed for computational efficiency.
To realize the multicanonical ensemble, we can perform a single-spin
flip with a flip rate of $\min\bigl(1, n(E)/n(E')\bigr)$.  However,
since the density of states $n(E)$ is not known beforehand, we have
proposed to use the count number $N(\sigma, \Delta E)$.  From the
broad-histogram equation, Eq.~(\ref{bheq}), the ratio of $n(\cdot)$ is
related to the ratio of $N(\cdot)$, we have
\begin{equation}
a(\sigma \to \sigma') = \min\left(1, 
{ \langle N(\sigma', E-E') \rangle_{E'}  \over \langle N(\sigma, E'-E)\rangle_E } \right).
\end{equation}
This is our flat-histogram flip rate.  Although the microcanonical
average $\langle N(\sigma, \Delta E)\rangle_E$ is also not available
before the simulation, it can be obtained approximately during a
simulation.  We use the instantaneous value and running average to
replace the exact microcanonical average.  Numerical tests have shown
that this procedure converges to the correct ensemble for sufficiently
long runs. For realizing truly a Markov chain, it is sufficient for a
two-pass algorithm.  The first pass is as before.  In the second pass,
we use a multicanonical sampling rate, using the density of states
determined from the first pass.

A variation of the algorithm is equal-hit algorithm which combines the
N-fold way method \cite{bortz-kalos-lebowitz} with a flip rate that
gives an extended ensemble that is uniform in probability of visiting
each new energy.  Reference~\cite{JSP} gives more extensive
discussions, as well as comparison with Wang-Landau method
\cite{wang-landau}.

\section{Some Results}

As there are more detailed balance equations among the transitions of
different energies than the number of energy levels, we determine the
density of states from the transition matrix by solving a
least-squares problem, or more generally, a nonlinear optimization
problem.  The optimization can be done either in the density of states
$n(E)$, or in the transition matrix elements $T(E \to E')$.  There are
a number of constraints that the transition matrix must satisfy.  The
trivial one is the normalization, $\sum_{\Delta E} T(E \to E + \Delta
E) = 1$.  There exists a rather interesting constraint, known as $TTT$
identity, as well:
\begin{equation}
T(E \to E') T(E' \to E'') T(E'' \to E) = 
T(E \to E'') T(E'' \to E') T(E' \to E).
\end{equation} 
These constraints complicate the optimization algorithms.

While any of those extended ensemble methods reduce their efficiency
as the system size increases, the accuracy of a two-pass
flat-histogram/multicanonical simulation is rather good for a given
fixed amount of CPU times \cite{JSP,binary}.  The method can also give
excellent result for large systems, such as a $256 \times 256$ Ising
lattice \cite{MCQMC}, using a parallelized version of the program.
The method is also applied to a lattice protein model, the HP model,
with good performance \cite{lee-wang}.

A possible measure of computational efficiency is through the
tunneling times.  The tunneling time is defined to be the number of
Monte Carlo steps in units of a lattice sweep ($N$ basic moves), for
system making a pass from the highest energy level to lowest level, or
vice versa.  For the two-dimensional Ising model, this tunneling time
diverges with system sizes according to $L^{2.8}$ which is worse than
standard random walk.  On the other hand, for spin-glasses with
complicated low-temperature free-energy landscape, the tunneling time
is much larger.  It is about $L^{4.7}$ \cite{zhan-lee-wang-physica-A}
in two dimensions and $L^{7.9}$ \cite{JSP} in three dimensions.
Another measure for spin glasses is given by the average first-passage
times.  It is defined as the average number of sweeps needed to reach
a ground state.  It is found in ref.~\cite{jpsj-okabe} that the
first-passage time diverges exponentially rather than according to a
power.  In any case, the equal-hit algorithm performs comparable to
`extremal optimization' \cite{EO} which is an optimization algorithm
inspired from self-organized criticality.

\section{Conclusion}

By collection the transition matrix, more information is obtained
about the system, giving more accurate results.  The effect of using
transition matrix is more dramatic for small systems.  Although the
transition matrix analysis of data can be used with any simulation
algorithms, extended-ensemble-based algorithms, such as flat histogram
algorithm, are excellent choices.  The efficiency of the
flat-histogram related algorithms has be studied.


\section*{Acknowledgments}
The author thanks J. Gubernatis for invitation to this conference
``The Monte Carlo method in the physical sciences: celebrating the
50th anniversary of the Metropolis algorithm'' at Los Alamos.  He would
also like to thank Robert H. Swendsen, Lik Wee Lee, Zhifang Zhan, and
Yutaka Okabe for collaborations on topics discussed here.





\begin{thebibliography}{10}

\bibitem{histo1} A. M. Ferrenberg and R. H. Swendsen, \textsl{%
Phys. Rev. Lett.} \textbf{61}, 2635 (1988).

\bibitem{histo2} A. M. Ferrenberg and R. H. Swendsen, \textsl{%
Phys. Rev. Lett.} \textbf{63}, 1195 (1989).

\bibitem{wang-tay-swendsen-PRL} J.-S. Wang, T. K. Tay, and R. H. Swendsen, 
\textsl{Phys. Rev. Lett.} \textbf{82}, 476 (1999);  J.-S. Wang, \textsl{Comp.
Phys. Commu.} \textbf{121-122}, 22 (1999).

\bibitem{Li-thesis} S.-T. Li, ``The transition matrix Monte Carlo  method,''
Ph.D. dissertation, Carnegie Mellon  University (1999), unpublished.

\bibitem{JSP} J.-S. Wang and R. H. Swendsen, \textsl{J. Stat. Phys.}
\textbf{106}, 245 (2002).

\bibitem{wang-europhys-B} J.-S. Wang, \textsl{Eur. Phys. J. B} 
\textbf{8}, 287 (1999); Physica A \textbf{281}, 147 (2000); 
in \textsl{`Computer Simulation Studies in Condensed-Matter Physics XIV'}, 
p. 113, Eds. D. P. Landau, S. P. Lewis, and H. B.  Sch\"uttler 
(Springer Verlag, Heidelberg, 2002).

\bibitem{swendsen-int} R. H. Swendsen, B. Diggs, J.-S. Wang, S.-T. Li,  C.
Genovese, J. B. Kadane, \textsl{Int. J. Mod. Phys. C} \textbf{10}, 1563
(1999).

\bibitem{wang-lee} J.-S. Wang and L. W. Lee, \textsl{Comp. Phys. Commu.} 
\textbf{127}, 131 (2000).

\bibitem{oliveira-brasil1} P. M. C. de Oliveira, T. J. P. Penna,  and H. J.
Herrmann, \textsl{Braz. J. Phys.} \textbf{26}, 677 (1996).

\bibitem{oliveira-europhys-b-2} P. M. C. de Oliveira, \textsl{Eur. Phys. J. B%
} \textbf{6}, 111 (1998); \textsl{Braz. J. Phys.} \textbf{30}, 766 (2000).

\bibitem{oliveira-broad} P. M. C. de Oliveira, T. J. P. Penna,  and H. J.
Herrmann, \textsl{Eur. Phys. J. B} \textbf{1}, 205 (1998);  P. M. C. de
Oliveira, \textsl{Braz. J. Phys.} \textbf{30}, 195 (2000).

\bibitem{berg-hansmann-europhys-b} B. A. Berg and U. H. E. Hansmann, \textsl{%
Eur. Phys. J. B} \textbf{6}, 395 (1998).

\bibitem{berg} B. A. Berg and T. Neuhaus, \textsl{Phys. Rev. Lett.} \textbf{%
68}, 9 (1992);  B. A. Berg, \textsl{Inter. J. Mod. Phys. C} \textbf{3}, 1083
(1992);  B. A. Berg, \textsl{Fields Inst. Commun.} \textbf{26}, 1 (2000).

\bibitem{bortz-kalos-lebowitz} A. B. Bortz, M. H. Kalos, J. L. Lebowitz, 
\textsl{J. Comput. Phys.} \textbf{17}, 10 (1975).

\bibitem{wang-landau} F. Wang and D. P. Landau, \textsl{Phys. Rev. Lett.}
\textbf{86}, 2050 (2001).

\bibitem{binary} J.-S. Wang, O. Kozan, and R. H. Swendsen, 
to appear in \textsl{`Computer Simulation Studies in Condensed Matter 
Physics XV,'} Eds.
D. P. Landau, S. P. Lewis, and H. B. Schuettler (Springer Verlag, Heidelberg, 2003). 
 
\bibitem{MCQMC} J.-S. Wang, in \textsl{`Monte Carlo and Quasi-Monte Carlo
Methods 2000,'} K.-T. Fang, F. J. Hickernell, and H. Niederreiter (Eds.),
p.141, Springer-Verlag, Berlin (2002).

\bibitem{lee-wang} L. W. Lee and J.-S. Wang, 
\textsl{Phys. Rev. E}, \textbf{64}, 056112  (2001).

\bibitem{zhan-lee-wang-physica-A} Z. F. Zhan, L. W. Lee, and J.-S. Wang, 
\textsl{Physica A} \textbf{285}, 239 (2000).

\bibitem{jpsj-okabe} J.-S. Wang and Y. Okabe, 
to appear, \textsl{J. Phys. Soc. Jpn.} \textbf{72}, June 2003. 

\bibitem{EO} S. Boettcher and A. G. Percus, \textsl{Phys. Rev. Lett.}
\textbf{86}, 5211 (2001). 

\end{thebibliography}

\end{document}